\documentclass[reprint,aps,pra,twocolumn,showpacs]{revtex4-1}
\usepackage{CJK}
\usepackage{bm,amsmath,amssymb,amsfonts}
\usepackage{color}
\usepackage{dcolumn}
\usepackage{graphicx}
\usepackage{float}
\usepackage{hyperref}
\hypersetup{colorlinks=true,citecolor=blue,urlcolor=blue,linkcolor=blue}
\usepackage{ulem}
\usepackage{times}
\usepackage{subfigure}

\begin{document}
\title{Quantum phase transition of the two-dimensional Rydberg atom array in an optical cavity}
\author{Gao-Qi An}
\affiliation{Department of Physics, and Chongqing Key Laboratory for Strongly Coupled Physics, Chongqing University, Chongqing, 401331, China}
\author{Tao Wang}
\thanks{corresponding author: tauwaang@cqu.edu.cn}
\affiliation{Department of Physics, and Chongqing Key Laboratory for Strongly Coupled Physics, Chongqing University, Chongqing, 401331, China}
\author{Xue-Feng Zhang}
\affiliation{Department of Physics, and Chongqing Key Laboratory for Strongly Coupled Physics, Chongqing University, Chongqing, 401331, China}
\begin{abstract}
We study the two-dimensional Rydberg atom array in an optical cavity with help of the meanfield theory and the large-scale quantum Monte Carlo simulations. The strong dipole-dipole interactions between Rydberg atoms can make the system exhibit the crystal structure, and the coupling between two-level atom and cavity photon mode can result in the formation of the polariton. The interplay between them provides a rich quantum phase diagram including the Mott, solid-1/2, superradiant and superradiant solid phases. As the two-order co-existed phase, the superradiant solid phase breaks both translational and U(1) symmetries. Based on both numerical and analytic results, we found the region of superradiant solid is much larger than one dimensional case \cite{zhang01}, so that it can be more easily observed in the experiment. Finally, we discuss how the energy gap of the Rydberg atom can affect the type of the quantum phase transition and the number of triple points.
\end{abstract}
\maketitle

\section{INTRODUCTION}
Introducing the strong interactions into the quantum simulator is the key topic, because it is vital for simulating and studying the quantum phase transition (QPT) of the strongly correlated system \cite{QS}. As one of the most possible candidate, the Rydberg atoms stay at high level state with large principle quantum number $n$, so that they possess two main advantages, long lifetime ($\approx100\mu$s at $n\approx50$) and strong dipole-dipole interaction \cite{Browaeys2020}. In order to simulate the quantum many-spin system, the Rydberg atoms are loaded into the optical lattice at first \cite{Bloch2012,Bloch2015}. However, in contrast to the small lattice spacing ($< 1\mu$m), the blockade radius of Rydberg atom is so large (typically $R_6>5\mu$m) that few can be excited to the Rydberg state \cite{Block,2010RvM}. Recently, due to the rapid development of optical tweezer arrays, the Rydberg atom can be trapped in each tweezer site with arbitrary geometry \cite{Scholl2021,Ebadi2021}. The programmable Rydberg atom array boosts the whole field, such as the gauge theory \cite{LGT1,LGT2}, quantum topological phase \cite{SPT,SL}, the non-equilibrium quantum many-body system \cite{MBS} and so on \cite{review_np}. However, the laser lights are taken as the classical light field, so the corresponding Hamiltonian is more like ``classic".

On the other hand, the quantized light field can be introduced by loading the ultra-cold atoms into an optical high-fineness cavity \cite{cavity_review}. Then, the interactions between atoms and photons are strongly amplified \cite{Baumann2010,science.1220314}, and some exotic quantum phases emerge, such as the ``supersolid" phase \cite{Landig2016} and superradiant phase \cite{whb}. Although, some of them belong to quantum few body systems, the theory and numerical simulation demonstrate the generalized scaling relation of QPT can still be detected \cite{rabi0,rabi1,rabi2,rabi3,rabi4}.

Inspired by the recent experimental progresses in both Rydberg atom array and cavity-QED, it is valuable to discuss the QPT of the Rydberg atom array in an optical cavity. In our previous study \cite{zhang01}, because of the interplay between photon-mediated interaction and dipole-dipole interaction, the superradiant solid (SRS) phase is found via large-scale numerical simulation. This quantum phase breaks both translational symmetry and U(1) symmetry, which is reminiscent of the ``supersolid" phase. However, possibly due to the strong fluctuation in low dimension, the parameter region of the SRS phase is very narrow. Considering the unavoidable influence, such as photon leaking, it is extremely hard to detect it in the real experiment.

\begin{figure}[t]
	\centering      
	\includegraphics[width=0.5\textwidth]{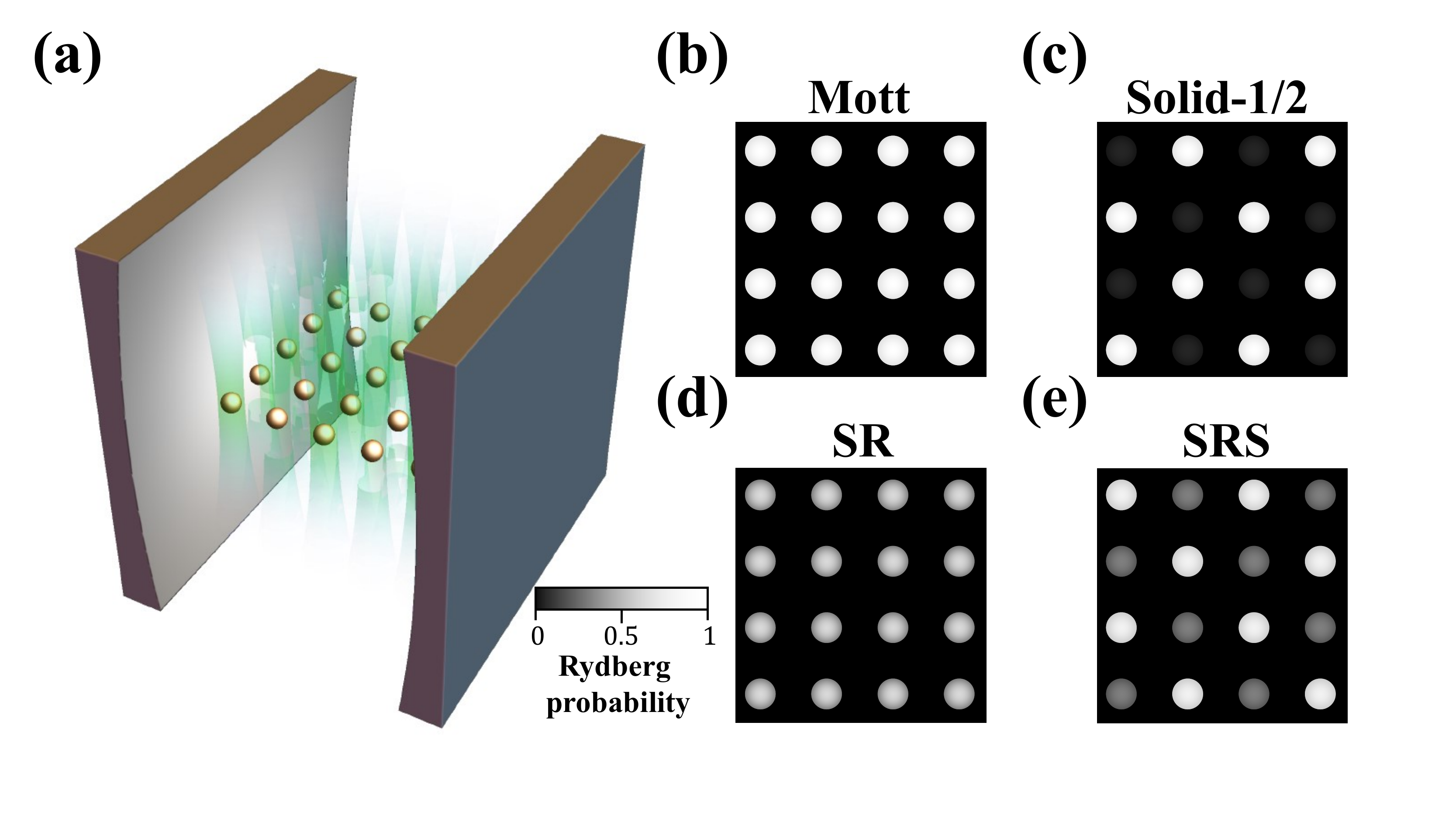} 
	\caption{ (a) A schematic of two-dimensional Rydberg atom array in a optical cavity. (b-e) The demonstrations of in situ Rydberg probability at different quantum phases in 4$\times$4 arrays.}
	\label{fig1}
\end{figure}

In this article, we study the QPT of the two-dimensional Rydberg atom array in a cavity shown in Fig.\ref{fig1}(a). By using both the analytic approaches and large-scale quantum Monte Carlo (QMC) simulation, we determine the quantum phase diagram composed with Mott, solid-1/2, superradiant (SR) and SRS phases. The configuration of these phases are illustrated in Fig.\ref{fig1}(b-e). In comparison with one-dimensional case \cite{zhang01}, the region of the SRS phase becomes broad. Meanwhile, the type of phase transition is analyzed in detail, especially the triple points.

The manuscript is organized as follows. In Sec. \ref{MODEL}, we discuss the model and its relation to both the Dicke model and the Ising model. In Sec. \ref{Meanfield Theory}, the mean field theory is borrowed to analyze the quantum phase transition and obtain the whole phase diagrams. In Sec. \ref{QUANTUM MONTE CARLO}, we implement the QMC method and compare the numerical results with the analytic results. Finally, in Sec. \ref{outlook}, we make the conclusion.

\newpage
\section{MODEL}
\label{MODEL}
In the experiment \cite{Browaeys2020}, the ultra-cold atom can be loaded into the defect-free tweezer arrays, and the Rydberg state can be excited via two-photon transitions. Then, if the transition between intermediate state $|i\rangle$ and the ground state $|g\rangle$ couples to a single quantized cavity mode, the whole system can be described by following Hamiltonian in the rotating wave approximation \cite{PhysRevA.82.053832,zhang01}:
\begin{eqnarray} 
\nonumber
H&=&\frac{g}{\sqrt{N}}\sum_{i=1}^{N}\left (b_{i}^{\dagger}a+h.c.\right )+C_{6}\sum_{\left \langle i,j \right \rangle}n_{i}n_{j}\\
&&-\Delta\sum_{i=1}^{N}n_i- \mu N_t, \label{E1}
\end{eqnarray}
where $a^{\dagger}$($a$) is the creation (annihilation) operator of photons, $b^{\dagger}_i$ ($b_i$) is the creation (annihilation) operator of Rydberg atom at site $i$, $n_{i}=b_{i}^{\dagger }b_{i}$ is the local Rydberg probability operator, $\Delta\ge0$ is the energy gap which can be changed by adjusting the laser detuning, $\mu$ is the chemical potential in the grand canonical ensemble, $g$ is the effective atom-photon coupling strength which is related to both cavity parameter and Rabi frequency of transition between Rydberg state $|e\rangle$ and intermediate state $|i\rangle$, $C_{6}$ denote the strength of the dipole-dipole interactions between Rydberg atoms \cite{PhysRevLett103185302}, and $\langle i,j \rangle$ represents only nearest neighbor interaction is considered. The maximum occupation number at Rydberg state in each tweezer site should be one, so $b$ should be treated as hard-core bosonic operator. It is obvious that the total density $N_t\equiv a^{\dagger}a+\sum_{i=1}^{N}n_i$ is a conserved quantity. Meanwhile, the model preserves the U(1) symmetry which is $a^{\dagger}\rightarrow a^{\dagger}e^{i\theta}$ and $b^{\dagger}\rightarrow b^{\dagger}e^{-i\theta}$. Notice that, the chemical potential $\mu$ should be $\le0$, otherwise the number of photons will diverge.

In the weak interaction limit $C_6\rightarrow0$, the model changes into the Dicke model within the rotating wave approximation \cite{PhysRevE066203,PhysRevA7831}. If the magnitude of the atom-photon coupling $g$ is small, the system is in the normal phase or Mott-0 phase in which all the atoms stay in the ground state. Then, when $g$ is large, the atom and photon can form the polariton, so that the system enters into the SR phase which breaks the U(1) symmetry (Fig.\ref{fig1}(d)). At critical point $g_{c}=\sqrt{|\mu(\mu+\Delta)|}$, a second-order QPT occurs.

In the strong interaction limit $g\rightarrow0$, the photon mode is decoupled with the Rydberg atoms. Then, the Hamiltonian can be reduced to Ising model by implementing the conventional mapping between hard-core boson and spin half operators $b^{\dagger}_i\rightarrow S^+_i$, $b_i\rightarrow S^-_i$ and $n_i\rightarrow S_i^z+1/2$. Then, the quantum phase diagram at zero temperature can be exactly obtained by calculating the energy of different configurations. When $\mu< -\Delta$, all atoms are at the ground state, i.e., the Mott-0 phase. After increasing $\mu$ to be larger but less than $4C_{6}-\Delta$, the atoms on one sublattice are excited to the Rydberg state (Fig.\ref{fig1}(c)), so the translational symmetry is spontaneously broken and the solid-1/2 phase (antiferromagnetic phase in spin language) is constructed. Then, continuously increasing $\mu$ until larger than $4C_{6}-\Delta$ results in that all the atoms are excited to the Rydberg state, and the system enters into the Mott-1 phase  (Fig.\ref{fig1}(b)).

In the intermediate region, the atom-photon coupling can provide photon-mediated long-range interaction, so that the SRS phase emerges accompanying with both U(1) and translational symmetries spontaneously broken. In comparison with previous work \cite{zhang01}, the Hamiltonian \ref{E1} is a hybrid 0d-2d quantum system, so the quantum fluctuation should be weaker in spirit of Mermin-Wagner theorem \cite{Mermin}. The effect of dimension should strongly change the whole quantum phase diagram including the QPT. Thus first, we prefer to study the model with the mean field theory.

\section{Meanfield Theory}
\label{Meanfield Theory}
In the spin wave theory, the first step is taking the quantum spin as the classical one, and then finding the orientation of spins with lowest energy. Such meanfield theory (or the semi-classical approximation) is also suitable for our model, because the Rydberg atom is a kind of quantum spin half as mentioned before. Meanwhile, since the coherent state is the ``most classical" quantum state of photons, we introduce following ansatz of wavefunction as before \cite{zhang01}:
\begin{eqnarray}
|\lambda ,\theta \rangle&=e^{\frac{\lambda \sqrt{N}a ^{\dagger }}{2}}\prod_i\left[\cos(\frac{\theta_i}{2})b_i^{\dagger}+\sin(\frac{\theta_i}{2})\right]|0\rangle, \label{E2}
\end{eqnarray}
in which $|0\rangle$ represents the vacuum state, $\lambda$ and $ \theta_{i}$ are the variational parameters of photons and Rydberg atoms. Because only the nearest neighbor repulsive interaction is kept, the translational symmetry breaking can only result in the Neel order or $(\pi,\pi)$ order. Thus, we set the variables $\theta_i$ are same in the same sublattice C or D, then the variational parameters are only left with $(\lambda,\theta_C,\theta_D)$. The energy per site of the ansatz Eq.\ref{E2} can be calculated as $E=\left \langle \lambda ,\theta |H|\lambda ,\theta  \right \rangle/N$ and
\begin{eqnarray}
4E&=&g\lambda (\sin \theta _{C}+ \sin \theta _{D}  )+2C_{6}\cos \theta _{C} \cos \theta _{D}-\mu\lambda ^{2}\nonumber\\
&&-( \mu+\Delta -2C_{6})( \cos \theta _{C}+ \cos \theta _{D})+E_0
\end{eqnarray}
where  $E_0=-2( \mu +\Delta - C_{6})$ is the energy constant.  

The ground state can be calculated by minimizing the energy per site $E$ with respect to the variational parameters $\lambda$, $\theta _{C}$ and $\theta _{D}$. The system has the sublattice symmetry $C\leftrightarrow D$, so we can set $0\le\theta_{D}\le\theta_{C}<2\pi$. The population of atom at the Rydberg state or the Rydberg probability can be evaluated by $\rho_i=(\cos(\theta_i)+1)/2$. When the density $\rho_C$ is not equal to $\rho_D$, it means that the translational symmetry is spontaneously broken. Besides, the energy is unchanged under the transformation $\lambda\rightarrow-\lambda$ and $(\theta _{C},\theta _{D})\rightarrow-(\theta _{C},\theta _{D})$, so the photon parameter is set to $\lambda\ge0$. Because the ansatz is the coherent state, the expectation of photon density is $|\lambda|^2$. Then, the nonzero $\lambda$ indicates the U(1) symmetry is spontaneously broken. Without loss of generality, we take $C_6$ as the unit of energy and set it to one.
 
\begin{figure}[t]
	\includegraphics[width=0.5\textwidth]{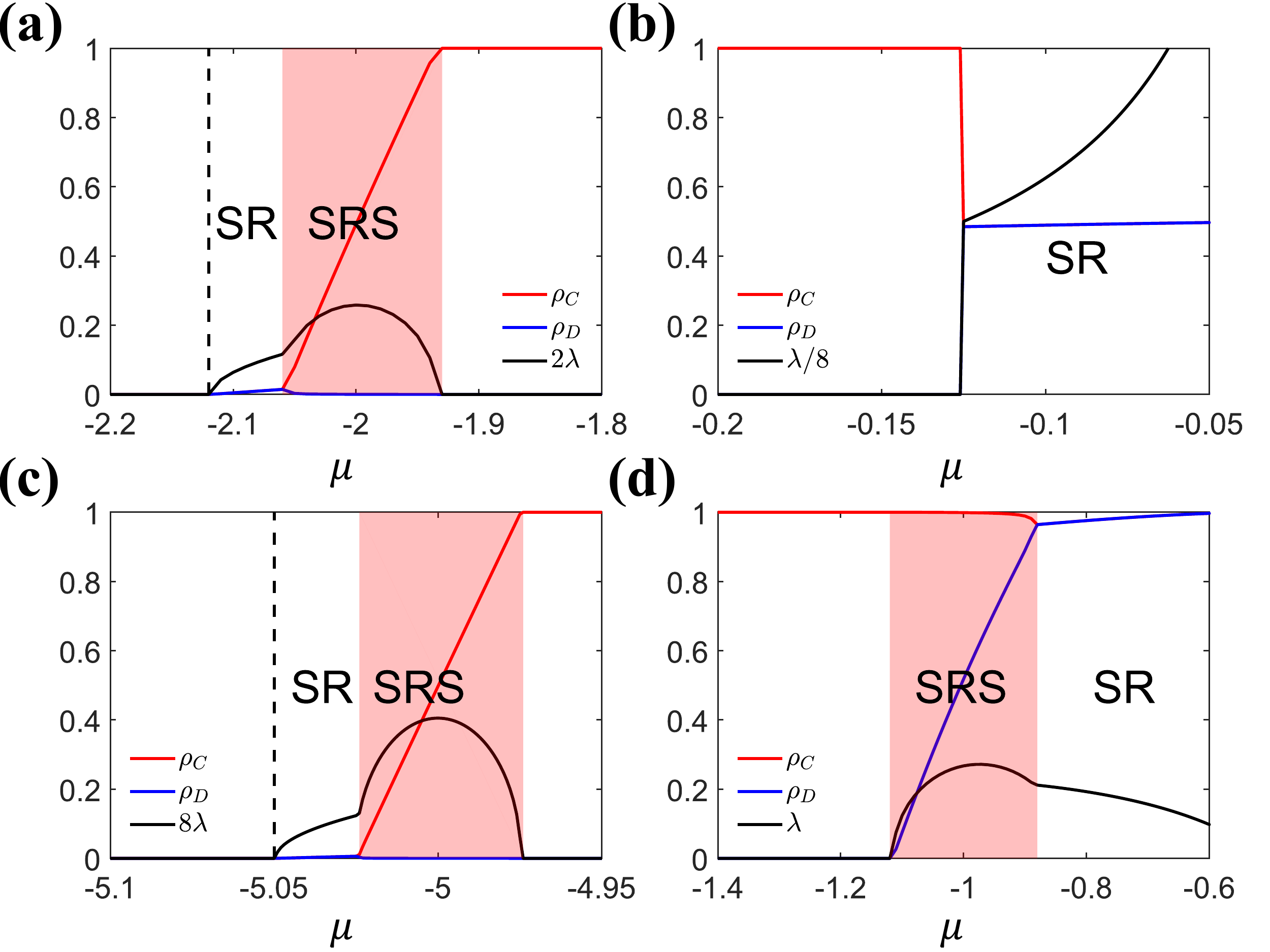} 
	\caption{Rydberg probabilities in different sublattices and the variational parameter of photon $\lambda$ are calculated at different chemical potential $\mu$ with help of meanfield theory. The atom-photon coupling strength is chosen $g=0.5$ with (a-b) $\Delta=2$ and (c-d) $\Delta=5$. The regions of the SRS phase is highlighted with red color.}
	\label{fig2}
\end{figure}

In Fig.\ref{fig2}, we show the Rydberg probabilities in different sublattices and photon parameter $\lambda$ at different $\Delta$. When $\mu$ is much smaller than $\Delta$, all the atoms stay at the ground state and form the Mott-0 phase. After increasing the chemical potential $\mu$ to a certain critical value, the photon number starts increasing and the Rydberg probabilities in both sublattice turns to be finite and equal. These phenomena demonstrate the quasi-particles polaritons are excited and break the U(1) symmetry, so that the SR phase emerges. Then, similar to the one-dimensional case \cite{zhang01}, the larger $\mu$ can not immediately drive the system into the solid-1/2 phase. Instead of it, the exotic SRS phase can be found between SR and solid-1/2 phase. From the order parameters in SRS phase shown in  Fig.\ref{fig2} (a,c), we can find both the photon numbers and Rydberg probabilities are finite. However, different from the SR phase, the Rydberg probabilities in both sublattices are different in SRS phase. It means that, as the two-order co-existed phase, the SRS phase breaks both U(1) and translational symmetries. In the strong coupling limit $g\ll\Delta$, SRS phase can be understood as the following picture: the atoms in one sublattice couple the photons and construct the polaritons \cite{zhang02}. While approaching the solid-1/2 phase, the photon density is dropping down to zero. In the solid-1/2, only the atoms in one sublattice are excited to the Rydberg state.

On the other side, when continuously increasing the chemical potential $\mu$, the existence of the SRS phase is highly relevant to the energy gap $\Delta$. When the gap is small, such as $\Delta=2$ in Fig.\ref{fig2}(b), the SRS phase is unstable, so that the QPT between SR and solid-1/2 phase is first-order. It can be obviously reflected by the jump of the Rydberg probabilities and the photon density. In comparison, the large energy gap (see Fig.\ref{fig2} (d)) can stabilize the SRS phase, in which the atoms of one sublattice are almost fully polarized and the another sublattice is occupied with the polaritons. Furthermore, the QPTs among SR, SRS and solid-1/2 phase are all second-order, which is the same as the situation at lower chemical potential.

\begin{figure}[t]
	\centering
	\includegraphics[width=0.5\textwidth]{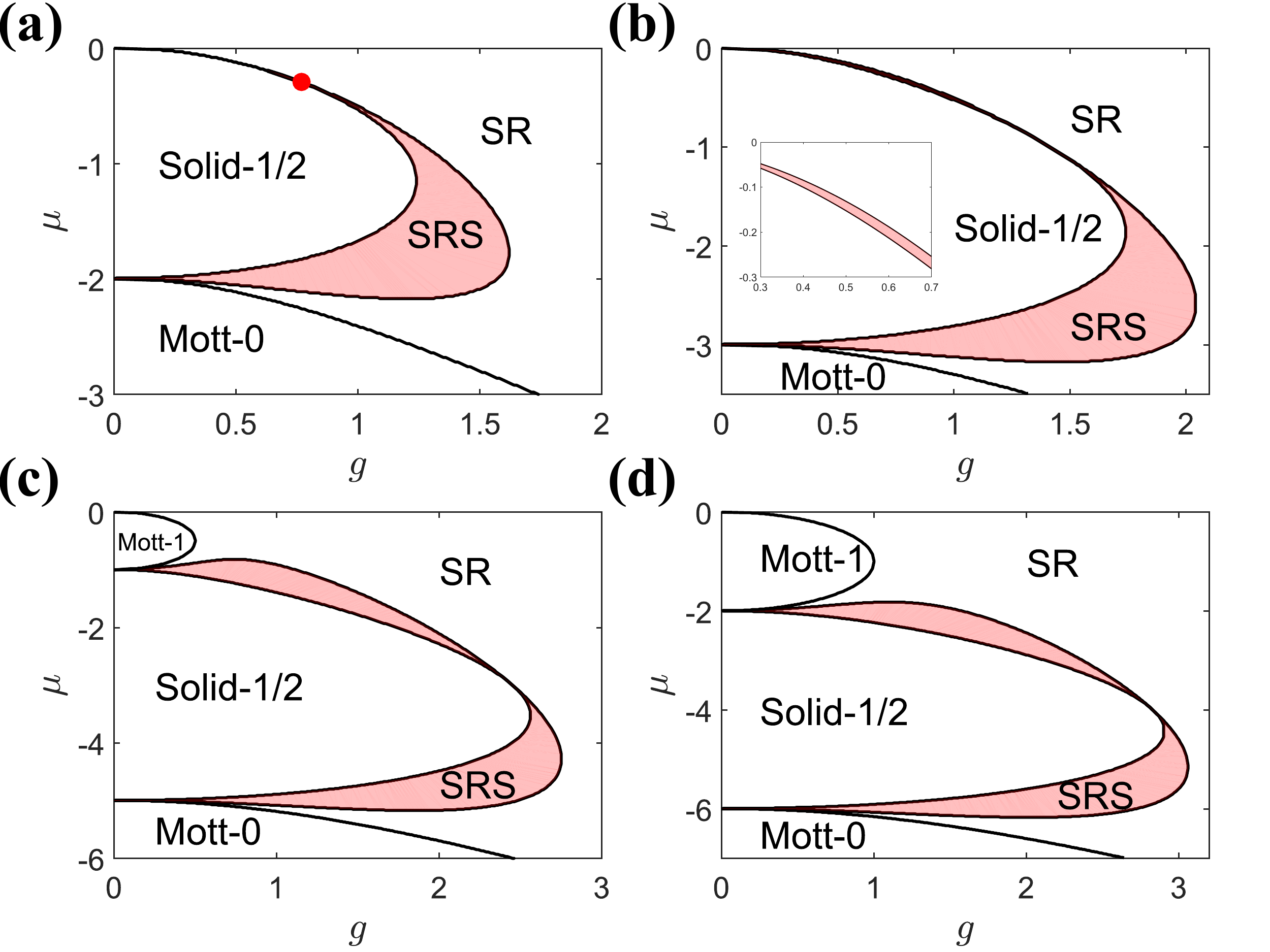}
	\caption{Quantum phase diagrams calculated by using the meanfield theory at different gap energy (a) $\Delta=2$, (b) $\Delta=3$, (c) $\Delta=5$, (d) $\Delta=6$. The region of SRS phase is marked with red color, and the red dot points out the triple point. Inset: the phase diagram at $\Delta=3$ after zooming in.}
	\label{fig3}
\end{figure}

The whole quantum phase diagram in $\mu-g$ panel with different gap energy $\Delta$ are drawn in Fig.\ref{fig3}. In both strong and weak coupling region, the solid-1/2, Mott and SR phases follow the analysis mentioned in Sec.\ref{MODEL}. In the intermediate region, the quantum phase diagram becomes rich. Apparently, the gap energy can stabilize the solid-1/2 and Mott phase, so that their region are enlarged while increasing $\Delta$. When gap energy is small (Fig.\ref{fig3}(a) at $\Delta=2$), the solid-1/2 phase has two paths of entering the SR phase: (i) a direct first-order phase transition through the upper phase boundary; (ii) two successive second-order phase transitions through the lower intermediate SRS phase. Meanwhile, there is a triple point among these phases (red dot in Fig.\ref{fig3}(a)). Then, while increasing $\Delta$, the upper SRS phase appears and its region is expanding. However, the upper and lower regions of SRS phase are not connected, and it hints there are two triple points jointed with first-order phase transition or one quadruple point. To address these questions and quantitatively study this system, we require a more accurate method, like the QMC simulation method.

\section{Quantum Monte Carlo Simulation}
\label{QUANTUM MONTE CARLO}
The numerical method we adopted is the stochastic series expansion method \cite{PhysRevB59R14157,PhysRevE66046701}. Different from the conventional algorithm, here the operator vertex includes three Rydberg sites and one photonic mode site. Meanwhile, the maximum occupation number of photons is adjusted to be always much higher than the possible photon number during the simulation, so there is no system error caused by the truncation. The algorithm in details can be found in supplementary materials of Ref.\cite{zhang01}. The inverse temperature is set to $\beta=1/T=100$ which is much lower than the other energy scales. The system size simulated is up to $20\times20=400$ sites with periodical boundary condition, and it is much larger than the real system ($\approx250$ \cite{Ebadi2021}). Usually, the distance between nearest neighbor sites in the experiment is around $5\mu$m, so the system size $20\times20$ implies the size of cavity has to be larger than $100\mu$m at least. However, it is not an easy task even with state-of-the-art techniques. 

\begin{figure}[t]
	\centering      
	\includegraphics[width=0.48\textwidth]{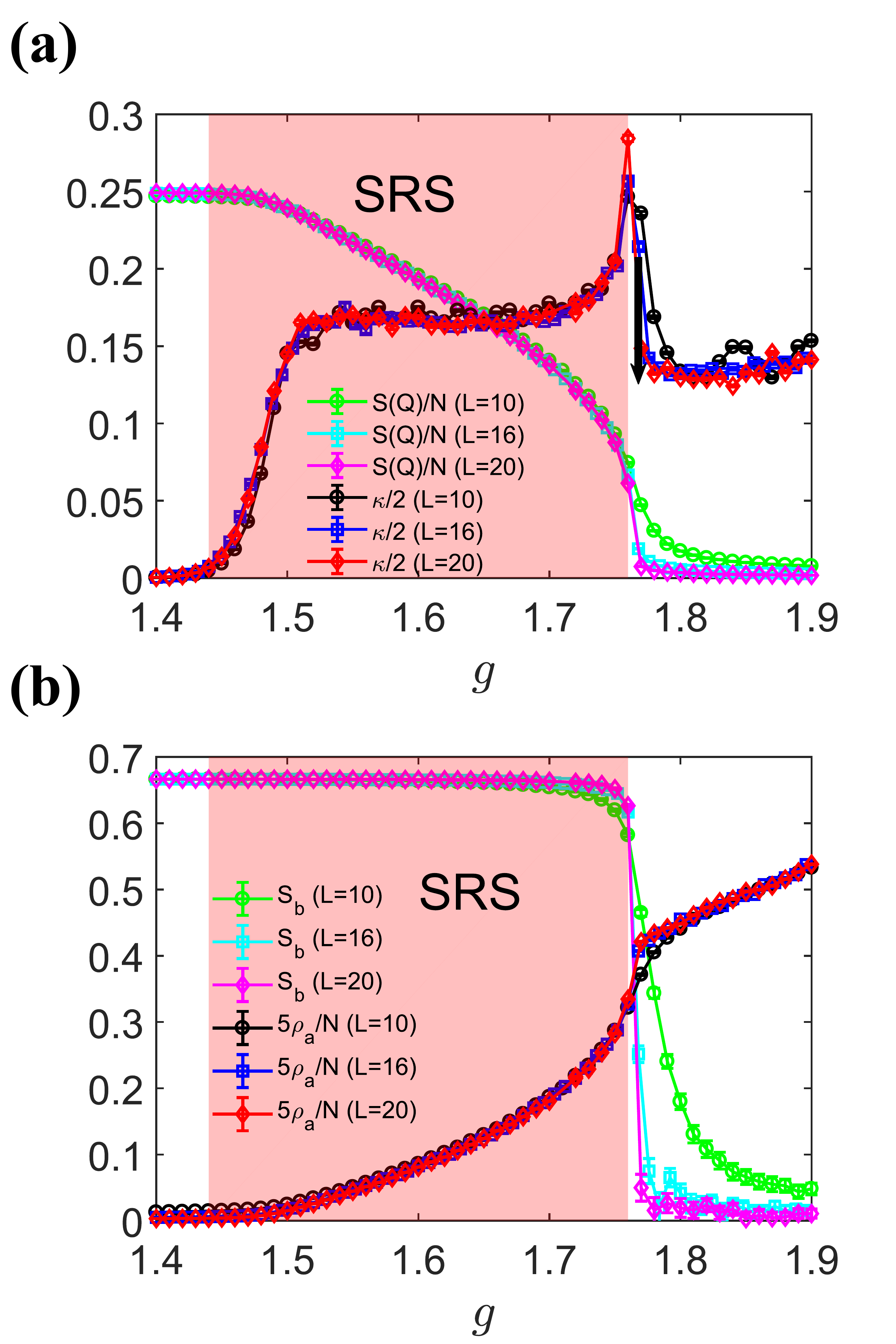}  
	\caption{(a)The structure factor $S(\textbf{Q})/N$ and the compressibility $\kappa$, (b) the Binder cumulant $S_b$ and the photon density $\rho_a$ at $\Delta=3$ and $\mu=-2.5$ with different system sizes. The region of SRS phase is marked with red color. The black arrow highlights the decreasing of $\kappa$ near the critical point.}
	\label{fig4}
\end{figure}
The possible quantum phases can divide into two categories: (1) the solid-1/2 and Mott phases are incompressible; (2) the SRS and SR phases are compressible. Thus, we introduce the compressibility $\kappa =N\beta ( \langle\rho ^{2}\rangle-\langle \rho \rangle^{2})$ to distinguish them, in which $\rho=\sum_{l=1}^{N}n_{l}/N$ is the average Rydberg probability. For the incompressible phases, Mott phases keep the integer filling and solid-1/2 phase corresponds to half filling. Turn to the compressible phase, the SRS phase further breaks the translational symmetry while the SR phase does not. Thus, the structure factor $S(\textbf{Q})/N=\langle | \sum_{l=1}^{N}n_{l}e^{i\textbf{Q}\cdot \textbf{R}_{l}} | ^{2} \rangle/N^{2}$  is taken as order parameter to characterize the solid order. In the solid-1/2 phase, same as the Neel phase in the magnetic materials, the corresponding $\textbf{Q}$ is equal to $(\pi,\pi)$. Then, the value of $S(\textbf{Q})/N$ of solid-1/2 is exact $1/4$ at $g=0$, because the Rydberg probability is equal to one in one sublattice and zero in the other. Meanwhile, it should be finite value in SRS phase and zero in the other phases.

The QPT between solid-1/2 and SRS phase can be clearly identified in Fig.\ref{fig4} at large chemical potential $\mu=-2.5$ with small gap energy $\Delta=3$. When the coupling $g$ is small, the compressibility $\kappa$ is zero and the structure factor $S(Q)/N$ is almost $1/4$, so the system is in the solid-1/2 phase. Meanwhile, the photon density $\rho_a=a^{\dagger}a$ is zero in solid-1/2 phase (see Fig.\ref{fig4}(b)). Then, same as the prediction of the meanfield theory, numerous polaritons are excited and prefer one sublattice, so we can observe both $S(Q)/N$ and $\kappa$ are finite in SRS phase. The photon density with different system sizes can be scaled to one line by dividing the number of Rydberg sites $N$ (see Fig.\ref{fig4}(b)), and it hints the polariton density is nearly unchanged while enlarging the system size. In other words, the average photon numbers for constructing the polariton is almost the same at different sizes. Furthermore, all the smooth curves of different observables demonstrate the QPT is the second-order.

Although the meanfield theory supports that the QPT between SRS and SR phases is continuous, on the contrary, the numerical results in other similar systems are different \cite{zhang01,zhang02}. In Fig.\ref{fig4}, we can find the QPT looks like continuous at small system size, such as $L=10$, but the structure factor suddenly jumps to $\approxeq0$ at the critical point $g_c=1.76$ for larger sizes. At the same time, such small discontinuity can also be observed from the photon density $\rho_a$. Most importantly, as shown in Fig.\ref{fig4}(a), the values of $\kappa$ at the peak increases along with the system size and tends to diverge. In contrast, it is decreasing at $g$ slightly larger than $g_c$ (black arrow highlights). To further verify the type of QPT, we calculate the Binder cumulant defined as $S_b=1-\frac{\langle S(\textbf{Q})^2 \rangle}{3\langle S(\textbf{Q}) \rangle^2}$ and plot it in Fig.\ref{fig4}(b). It is zero in SR phase, and about 2/3 in the ordered phase (both solid-1/2 and SRS break the translational symmetry). When $g$ is closed to the critical point $g_c$, the Binder cumulant also presents a clear jump at larger size, and it indicates the QPT is the first-order.

\begin{figure}[t]
	\centering      
	\includegraphics[width=0.5\textwidth]{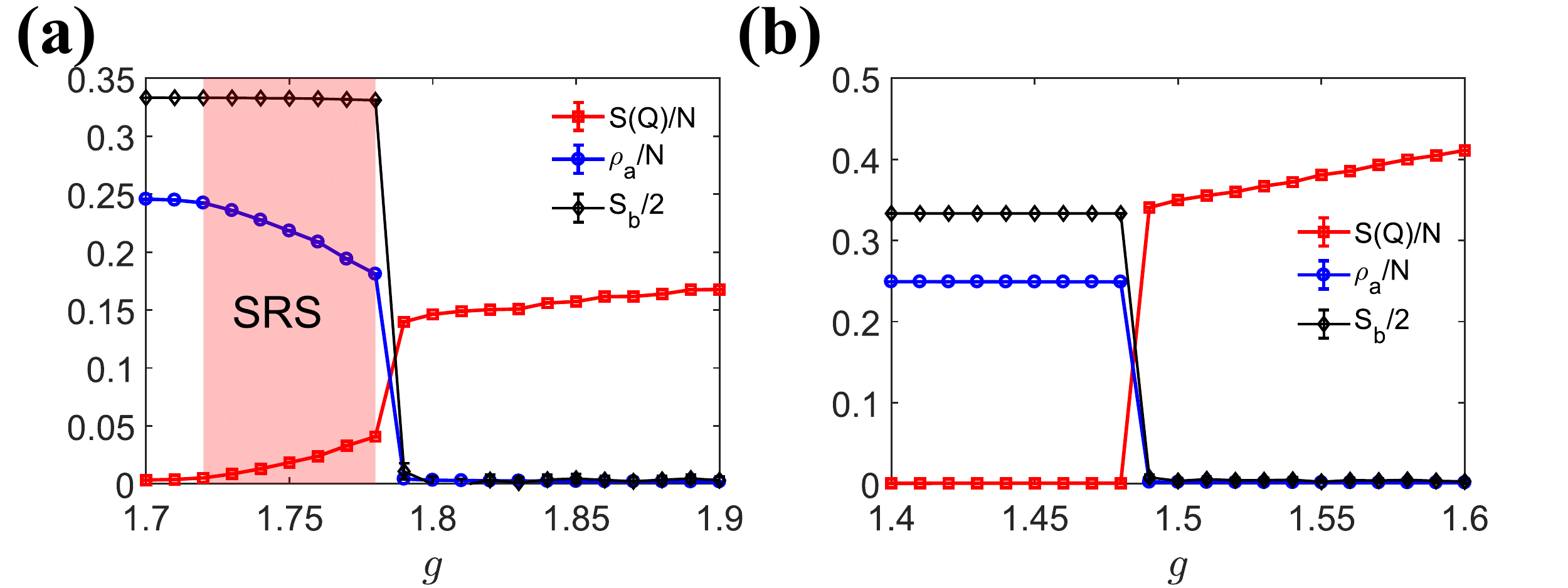}  
	\caption{The structure factor $S(\textbf{Q})/N$, its Binder cumulant $S_b$ and the photon density $\rho_a$ at $\Delta=3$ with system size equal to $L=20$. The chemical potential is set to (a) $\mu=-2.1$ and (b) $\mu=-1.2$. The region of SRS phase is marked with red color.}
	\label{fig5}
\end{figure}

When the chemical potential $\mu$ becomes smaller, as shown in Fig.\ref{fig5}(a), the region of the SRS phase shrinks. Meanwhile, the first-order phase transition between SRS and SR phase turn to be more apparent, and we can observe obvious jumps of all the observables. However, different from the predication of meanfield theory at $\Delta=3$, the QPT between solid-1/2 and SR phase turns to be direct first-order in Fig.\ref{fig5}(b). Indeed, such phenomenon also exists for the results of meanfield theory at smaller $\Delta$, and it reflects the meanfield theory underestimates the quantum fluctuations as usual. 

\begin{figure}[t]
	\centering
	\includegraphics[width=0.48\textwidth]{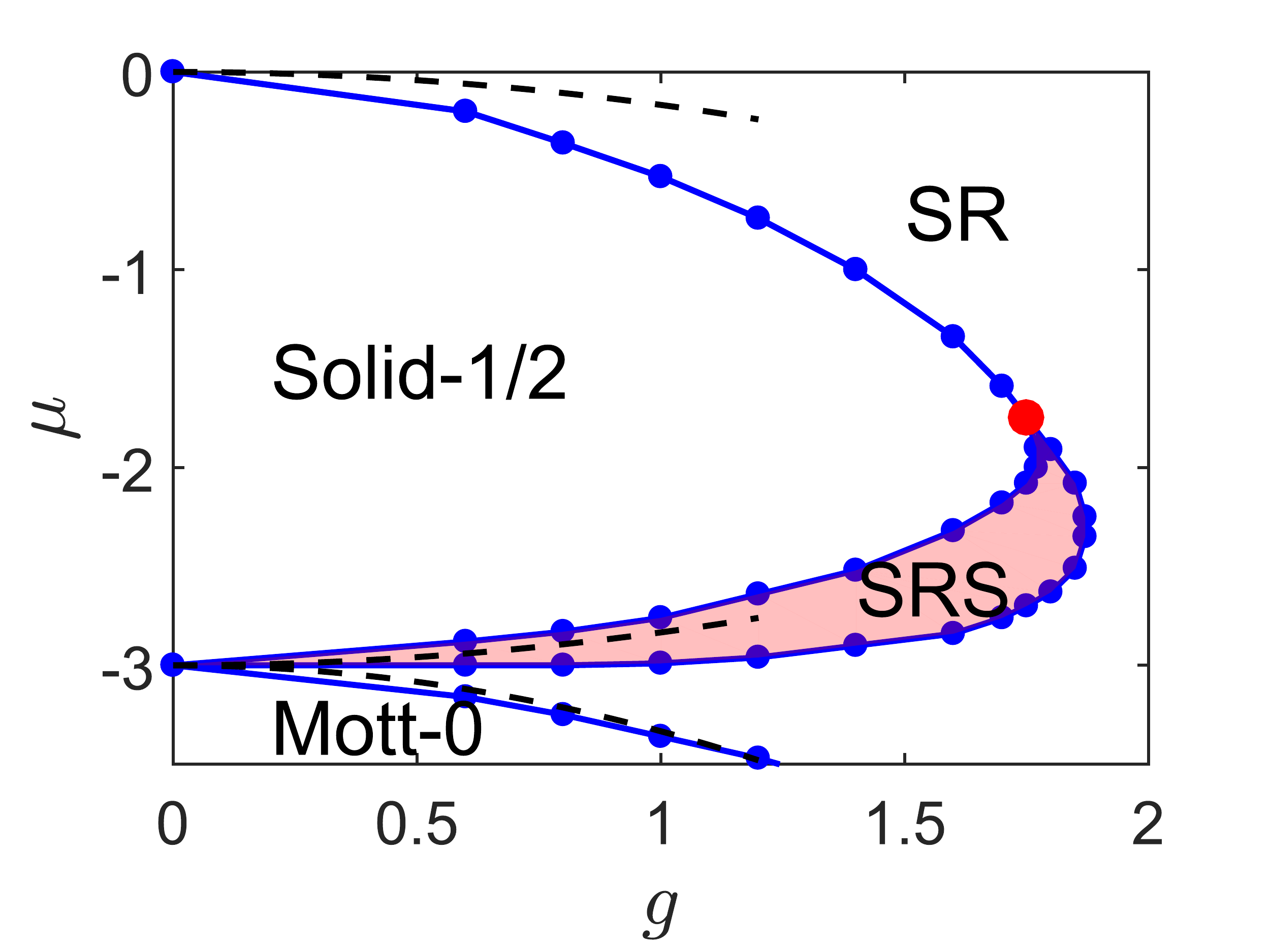}
	\caption{The quantum phase diagram at $\Delta =3$ and $L=10$. The blue dot lines are the phase boundaries got by using the QMC simulations, and the dashed lines are analytic results of strong coupling expansion method. The region of SRS phase is shaded with red color, and the red dot marks the triple point.}
	\label{fig6}
\end{figure}
In the real system, the number of tweezer sites is around $\approx$200. Thus, instead of tediously achieving the phase diagram in the thermodynamic limit, it is more practical of obtaining the ``finite-size phase diagram". Here, the system size of phase diagram is $N=100$. The phase boundary of incompressible phase is set to the value at which the compressibility $\kappa$ is just larger than $10^{-3}$. Meanwhile, the finite-size phase boundary between SRS and SR phase is found by the singularity of $\kappa$. From the quantum phase diagram at $\Delta=3$ in Fig.\ref{fig6}, we can find the region of SRS phase is slightly smaller than the meanfield phase diagram. It indicates the system behave more ``classically". In comparison with the one-dimensional case, the SRS phase is more stable, so that it will be more accessible in the real experiment. Meanwhile, it should be also possible to detect the triple point among SRS, SR, and solid-1/2 phases.

On the other hand, the critical lines of the second-order QPT of the incompressible phases can also be calculated via the perturbation theory, which is also known as the strong coupling expansion (SCE) method \cite{PhysRevB532691,PhysRevB8245107}. Several works have demonstrate it can provide very accurate results comparable to the numerical results. The phase boundary of Mott-0 phase can be calculated by comparing the vacuum energy with perturbative energy of one polariton excited state, and the second-order results is $\mu=-\Delta-g^{2}/\Delta$. The QPT from the solid-1/2 to SRS phases is the second-order, but the upper and lower critical lines should be discussed, separately. The lower one is caused by the `hole-like' excitations. When chemical potential is smaller than the critical line, one atom at Rydberg state can go back to the ground state and form a `hole'. Then, as the quasi-particle, the polariton can also be composed with the `hole' and photon. Thus, the energy difference between single `hole-polariton' excited state and solid-1/2 phase can give the second-order lower critical line $\mu=-\Delta+g^{2}/2\Delta $. For the upper boundary at $\Delta<4$, the first excited state is with additional photon, so the critical line is $\mu=-g^{2}/2\Delta$. 

We compare the SCE results with the numerical result in Fig.\ref{fig6}, and the critical lines of the Mott-0 phase quantitatively match well, so do the lower boundaries of QPT between SRS and Solid-1/2 phases. However, the upper boundary shows a large discrepancy, because the melting of solid-1/2 phase becomes the first-order. Inspired by both numerical and analytic results, we find it is hard to produce the upper SRS phase when $\Delta<4C_6$. One possible mechanism may be understood as follows: it is hard to excite the atom to the Rydberg state in solid-1/2 phase when $\Delta<4C_6$, so it is also difficult to construct the polaritons even with photons in the optical cavity.
\begin{figure}[t]      
	\includegraphics[width=0.48\textwidth]{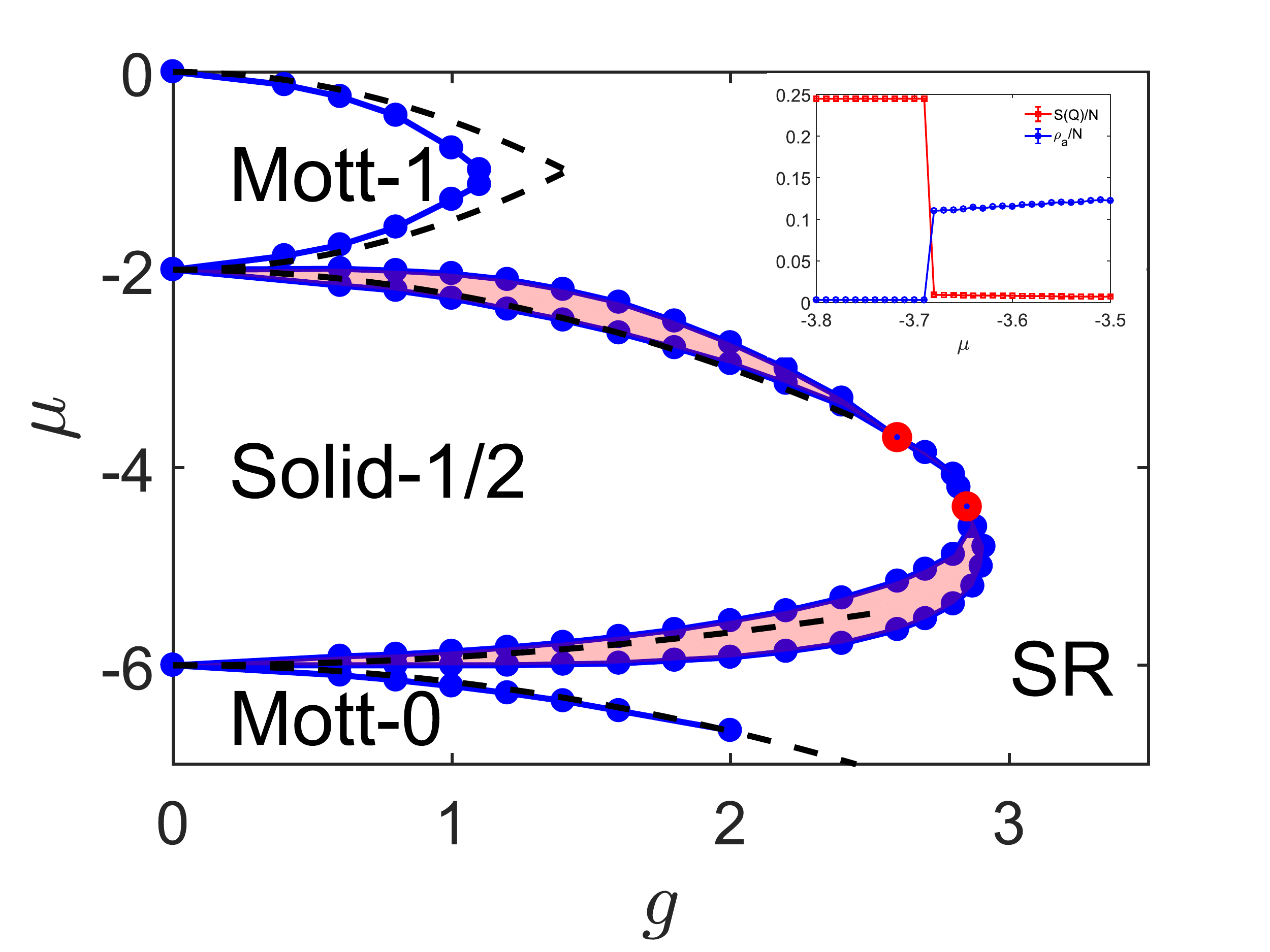} 
	\caption{The phase diagram at $\Delta=6$ and $L=10$. The blue dot lines are the phase boundaries got by using the QMC simulations, and the dashed lines are analytic results of strong coupling expansion method. The region of SRS phase is shaded with red color, and the red dots mark the triple points. Inset: the structure factor and photon density for different $\mu$ at $g=2.6$.} \label{fig7} 
\end{figure}

At last, we simulate the QPT at large gap energy $\Delta=6$, and the quantum phase diagram is shown in Fig.\ref{fig7}. The Mott-1 phase appears at high chemical potential. In Mott-1 phase, all the atoms stay at the Rydberg state, and the upper and lower boundaries can be obtained by calculating the perturbative energies of single `particle-polariton' or `hole-polariton' excited state. Then, the SCE method gives the second-order upper critical line $\mu=-g^{2}/( \Delta -4C_{6})$ and lower one $\mu =-(\Delta-4C_{6})+g^{2}/( \Delta -4C_{6})$. In the Fig.\ref{fig7}, we can find the critical lines of Mott-1 phase from both QMC and SCE methods match well at small $g$, and seem to be symmetric along $\mu=-1$ which may correspond to the hidden particle-hole symmetry. Furthermore, because the gap energy $\Delta$ is larger than $4C_6$, the SRS phase revives with `particle-polariton' excitation. Then, the second-order upper critical line of QPT between solid-1/2 and SRS phase changes to  $\mu=-(\Delta-4C_{6})-g^{2}/2( \Delta -4C_{6} )$. In the Fig.\ref{fig7}, all the critical lines from SCE method are very close to the numerical results at small $g$, and it demonstrates the physical mechanism of `polariton excitation driving QPT' is appropriate. Finally, between the SRS phases, there are two triple points linked with first-order phase transition (see the inset of Fig.\ref{fig7}), which is also found in the one-dimensional system \cite{zhang01}.

\section{Conclusion and Outlook}
\label{outlook}
In conclusion, borrowing the meanfield theory, strong coupling expansion method and larger scale quantum Monte Carlo simulation, we obtain the quantum phase diagram of Rydberg atoms trapped in square tweezer arrays inside an optical cavity. Not only the Mott phase, solid-1/2 and SR phase are observed, most importantly, the SRS phase which breaking both symmetries are well analyzed. The transition between SR and solid-1/2 phases can be the direct first-order, or the two successive second-order QPT through the intermediate SRS phase. The existence of the upper SRS phase is highly reverent to the gap energy, so is the triple points among SRS, SR and solid-1/2 phase. The phase transition between SRS and SR phase is found to be first-order, but we still can not rule out the continuous one in the real system due to the size or edge effect. Considering the rapid progress have been made in the fields of cavity and tweezer\cite{Scholl2021,Ebadi2021,Baumann2010,science.1220314}, our work will pave a way for the experiment in the near future and certainly provide benchmark for it at the same time.

\section{ACKNOWLEDGMENTS}
X.-F. Z. is thankful for valuable discussions with Qin Sun. This work is supported by
the National Science Foundation of China under Grants  No. 11874094 and No.12147102, Fundamental Research Funds for the Central Universities Grant No. 2021CDJZYJH-003.

\bibliographystyle{apsrev4-1}
\bibliography{referen}

\end{document}